\documentclass[12pt]{article}
\usepackage{amsmath, amssymb,graphicx}

\newwrite\ffile\global\newcount\figno \global\figno=1

\def\writedef#1{}

\input epsf
\def\figin{\epsfcheck\figin}\def\figins{\epsfcheck\figins}
\def\epsfcheck{\ifx\epsfbox\UnDeFiNeD
\message{(NO epsf.tex, FIGURES WILL BE IGNORED)}
\gdef\figin##1{\vskip2in}\gdef\figins##1{\hskip.5in}
\else\message{(FIGURES WILL BE INCLUDED)}%
\gdef\figin##1{##1}\gdef\figins##1{##1}\fi}

\def\figinsert{}
\def\ifig#1#2#3{\xdef#1{fig.~\the\figno}
\writedef{#1\leftbracket fig.\noexpand~\the\figno}%
\figinsert\figin{\centerline{#3}}\medskip\centerline{\vbox{\baselineskip12pt
\advance\hsize by -1truein\center\footnotesize{  Fig.~\the\figno.} #2}}
\bigskip\endinsert\global\advance\figno by1}
\def\endinsert{}

\usepackage{amssymb}
\usepackage{graphics}
\begin{document}
\baselineskip 18pt
\newcommand{\Tr}{\mbox{Tr\,}}
\newcommand{\beq}{\begin{equation}}
\newcommand{\eeq}{\end{equation}}
\newcommand{\bea}{\begin{eqnarray}}
\newcommand{\eea}[1]{\label{#1}\end{eqnarray}}
\renewcommand{\Re}{\mbox{Re}\,}
\renewcommand{\Im}{\mbox{Im}\,}

\def\N{{\cal N}}
\def\one{\hbox{1\kern-.8mm l}}


\thispagestyle{empty}
\renewcommand{\thefootnote}{\fnsymbol{footnote}}

{\hfill \parbox{4cm}{
        SHEP-07-18 \\
}}

\bigskip

\begin{center} \noindent \Large \bf
Quark Mass in the Sakai-Sugimoto Model of Chiral Symmetry Breaking
\end{center}

\bigskip\bigskip\bigskip

\centerline{ \normalsize \bf Nick Evans and Ed Threlfall
\footnote[1]{\noindent \tt
 evans@phys.soton.ac.uk, ejt@phys.soton.ac.uk
} }

\bigskip
\bigskip\bigskip

\centerline{ \it School of Physics \& Astronomy} \centerline{ \it
Southampton University} \centerline{\it  Southampton, S017 1BJ }
\centerline{ \it United Kingdom}
\bigskip

\bigskip\bigskip

\renewcommand{\thefootnote}{\arabic{footnote}}

\centerline{\bf \small Abstract}
\medskip

{\small \noindent We re-analyze $D8$ brane embeddings in the
geometry of a $D4$ brane wrapped on a circle that describe chiral
symmetry breaking in a strongly coupled non-supersymmetric gauge
theory. We argue that if the holographic fields are correctly
interpreted, the original embeddings describe a complex quark mass
and condensate in the theory. We show that in this interpretation
when a quark mass is present there is a massive pseudo Goldstone
boson (pion). A previously identified massless fluctuation is, we
argue, not a physical state in the field theory. We also determine
the behaviour of the quark condensate as a function of the quark
mass. }

\newpage


\section{Introduction}

A holographic model of chiral symmetry breaking in a strongly
coupled gauge theory has been recently proposed by Sakai and
Sugimoto \cite{Sakai:2004cn,Sakai:2005yt} (alternative string
realizations of QCD-like chiral symmetry breaking can be found in
\cite{Babington:2003vm}-\cite{Casero:2007ae}). Their model
consists of probe $D8/\bar{D8}$ branes in the geometry of a stack
of $N$ $D4$ branes wrapped on a circle.

The basic set up, for the simplest configuration, is shown on the
left side of Figure 1 - perturbatively the $D8$ and $\bar{D8}$ lie
in the non-compact dimensions of the space but are point like and
maximally separated on the wrapped circle. They intercept the $D4$
brane wrapped on the circle at the origin. Strings stretching
between the $D8$s and $D4$ provide massless chiral quark fields in
the field theory and the SU($N_f$) gauge symmetry of the $D8$ and
$\bar{D8}$ branes' world volumes correspond to the chiral symmetry
in the gauge theory.

When the full $D4$ brane geometry, with an induced horizon a
distance $u_{KK}$ away from the $D4$, is included though, the $D8$
and $\bar{D8}$ brane prefer to combine as shown by the curve
reaching in to the horizon on the right hand side of Figure 1.
There is a non-zero separation between the $D4$ and $D8$ branes
suggesting the dynamical generation of a quark mass ($\Sigma_q = T
u_{KK}$). The separate symmetries of the $D8$ and $\bar{D8}$ are
broken to a single vector symmetry on the combined world volume.
The model therefore encodes chiral symmetry breaking in the
pattern of QCD.

For this configuration which brings the $D4$ and $D8$ closest
(presumably corresponding to a massless initial quark), the
spectrum of fluctuations of the model, which correspond to bound
states in the gauge theory, has been well explored in the
literature \cite{Sakai:2004cn,Sakai:2005yt,fermions}.  A massless
Goldstone field (essentially the pion) exists in the spectrum of
the gauge field on the $D8$ brane.

However, we believe there is some confusion in the literature (eg
in
\cite{Sakai:2004cn}\cite{Antonyan:2006vw}\cite{Antonyan:2006pg}
\cite{Casero:2007ae}-\cite{Hashimoto:2007fa})
about how to interpret the remainder of the embedding solutions
shown on the right in Figure 1. For these solutions the minimum
$D4$ $D8$ separation increases which suggests that a bare quark
mass is being included. A naive analysis \cite{Sakai:2004cn} seems
to show though that the massless pion remains for these
configurations. There is a growing mythology that one can not
introduce a quark mass simply in this model and that the quark
condensate is not fixed by the dynamics.

In this letter we wish to re-analyze these `massive' embeddings
and argue that they do in fact correspond to the presence of a
bare quark mass and that there is no massless pion after all. We
point out that in the  weak coupling string construction there is 
generically an instability to the D8 and $\bar{D8}$ combining
and breaking chiral symmetry. The symmetry breaking can not be 
solely dynamically driven therefore (except for the specific case of
massless quarks when the D8 and $\bar{D8}$ are at anti-podal points 
on the circle).
In the strong coupling holographic description we will argue that the
quark condensate and mass are described by a single, complex
holographic field which we will identify. It is crucial to
correctly identify the quark mass and condensate within this
holographic field. We will show for example that the putative
massless pion in these embeddings is in fact a spurious field
corresponding to allowing the phase of the quark mass to vary
along with the condensate's phase. This is not a physical mode in
the gauge theory.

\begin{centering}
\begin{figure*}
\begin{centering}
\includegraphics[width=60mm]{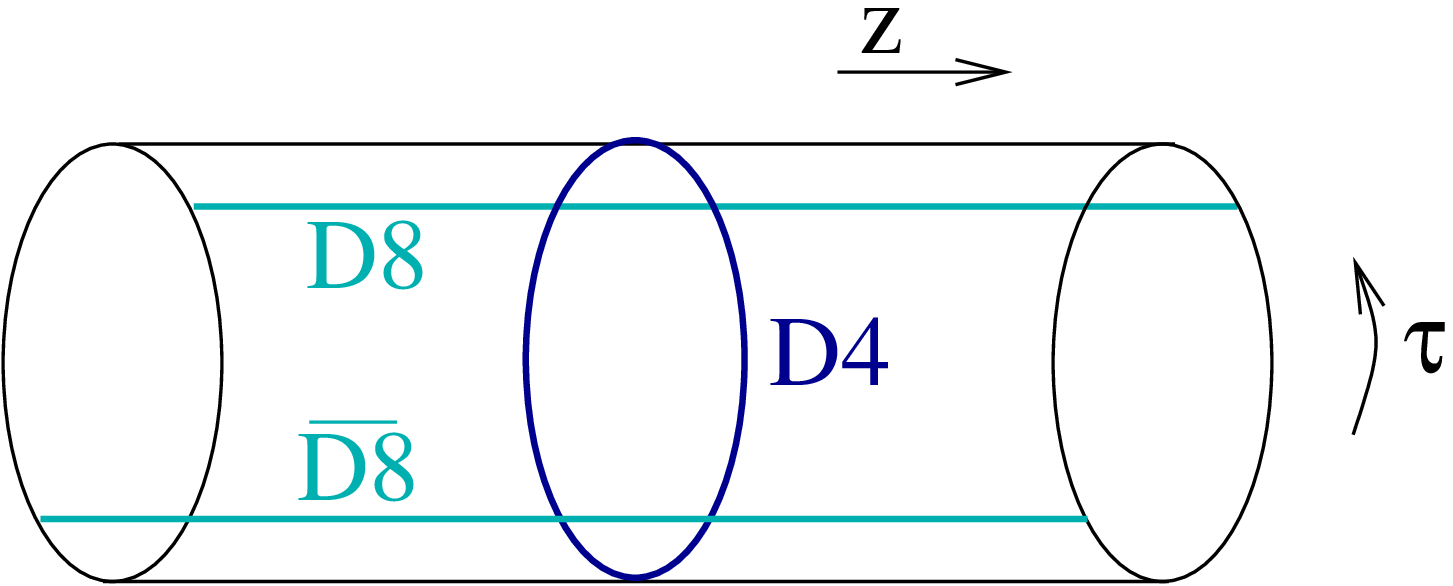} \hspace{2cm}
 \includegraphics[width=60mm]{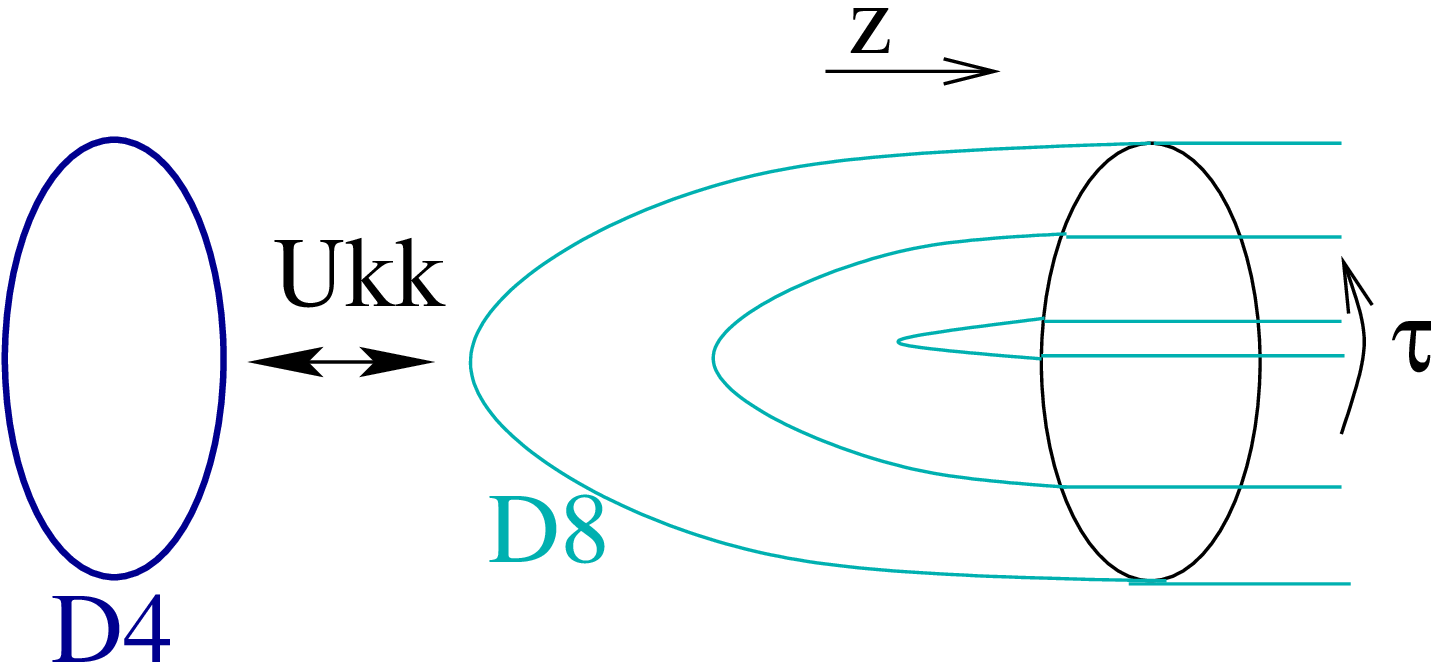}
\par
\end{centering}
\caption{Sketch of some D8-brane embeddings in the $z,\tau$-plane
both in the perturbative string construction and the gravity
dual.}
\end{figure*}
\par
\end{centering}

\section{The Sakai-Sugimoto Model - Perturbatively}

Let us begin by considering the possible $D8$ and $\bar{D8}$
configurations perturbatively in the Sakai-Sugimoto model (ie in
the left hand image of Figure 1). We will argue that for
generic separations in $\tau$ the picture in Figure 1 is not 
the perturbative ground state and is therefore misleading. The
straight D8 configurations do minimize the D8s' world volumes but
they do not take into account their mutual interactions (note
formally one might think these were down in a large $N$ expansion 
but the presence of a quark mass is independent of the gauge dynamics and 
so one must include the interactions).

The crucial observation is that
the system, which is non-supersymmetric, is unstable to the D8 and
$\bar{D8}$ coming together and annihilating. The set up is 
equivalent to placing an electron and a positron on a circle where their
attraction will bring them together to annihilate. The joining of
the D8 and $\bar{D8}$ corresponds to chiral symmetry breaking.

To be definite about the field theory one should consider at
least meta-stable configurations. There are two simple
set ups. Firstly if the branes lie at antipodal points on the
$S^1$, as discussed in the introduction, one would expect the
configuration to be static (but still unstable) - here we will
generate massless, chiral quarks. On the other hand if we lie the
$D8$ and $\bar{D8}$ on top of each other they will annihilate and
there will be no quarks - this is only achieved in the field theory 
at weak coupling by
including an infinite quark mass.

For configurations between these two extremes we can try to place
the $D8$ and $\bar{D8}$ at separate values of $\tau$ and we might
think this would interpolate between the massless and massive
limit. The desire of the $D8$ and $\bar{D8}$ to annihilate
suggests an inherent chiral symmetry breaking parameter is present.
One would expect though that they would come together all along their
length and completely annihilate.
The mass term appears to want to grow and remove the quarks from the
theory. This presumably reflects the fact that if  one makes the mass 
term in a gauge theory a dynamical scalar then it can minimize 
the vacuum energy 
by removing the quarks from the theory. 

When we look at the gravity dual of the field theory at strong
coupling we will find that there are configurations  where
asymptotically the $D8$ and $\bar{D8}$ do lie at non-antipodal
points (the right hand side of Figure 1). These configurations
interpolate between the theory with massless quarks and the theory
with no quarks - we think it entirely natural to associate these
with the presence of a quark mass and will argue further for this
interpretation below. Possibly at strong coupling these
configurations are stabilized because the vacuum energy can be
lowered by the formation of the quark condensate meaning the
theory ``likes" to have quarks present.

\section{The Sakai-Sugimoto Model - Strong Coupling}

The metric for the space around a stack of $D4$ branes in standard
coordinates is

\beq ds^2 = \left ( \frac{u}{R} \right )^{\frac{3}{2}} \left  (
dx_4^2 +f(u) d\tau^2 \right ) + \left ( \frac{R}{u} \right
)^{\frac{3}{2}} \left ( \frac{du^2}{f(u)} +u^2 d \Omega_4^2 \right
) \eeq

With $f(u) \equiv 1- \left ( \frac{u_{KK}}{u} \right )^3$.   There
is a nonzero four-form flux (not important for this analysis) and
a dilaton $e^{-\phi}=g_s \left (\frac{u}{R} \right
)^{-\frac{3}{4}}$.

Note the coordinate $\tau$ is periodic with the period given  by
$\delta \tau= \frac{4 \pi}{3}
\frac{R^{\frac{3}{2}}}{u_{KK}^{\frac{1}{2}}}$ forming a $S^1$
which is wrapped by the $D4$-branes.  This compactification is
necessary in order to make the spacetime smooth and complete.
There is a horizon at $u=u_{KK}$ (where the radius of the $S^1$
$\rightarrow 0$) which means the co-ordinate $u$ is restricted to
the range $[u_{KK}, \infty]$.

We will change variables to the radial coordinate $z$ where
$1+z^2 = \left ( \frac{u}{u_{KK}} \right )^3$ so the geometry becomes

\beq \begin{array}{ccl} ds^2 &  = & \left ( \frac{u_{KK}}{R}
\right )^{\frac{3}{2}} \left (\sqrt{1+z^2} dx_4^2 +
\frac{z^2}{\sqrt{1+z^2}} \; d \tau^2 \right )\\
&& \left. \right. \hspace{2cm} +\left ( \frac{R}{u_{KK}} \right
)^{\frac{3}{2}} u_{KK}^2 \left (\frac{4}{9} (1+z^2)^{-\frac{5}{6}}
\; dz^2+ (1+z^2)^{\frac{1}{6}} \; d \Omega_4^2 \right )
\end{array}\eeq

We can find the embeddings of a probe $D8$-brane in the above
background.   These form a family of curves in the
($z$,$\;\tau$)-plane which we parameterize as $\tau(z)$.  The
Dirac-Born-Infeld (DBI) action for the embedding is

\beq \mathcal{S}_{DBI} = \int_{D8} d^8 \zeta \; e^{-\phi}  \sqrt{-
Det[\mathcal{P}\left ( g_{ab} \right )]} \eeq

This gives

\beq \label{taction} \begin{array}{ccl} \mathcal{S}_{DBI}&  = &
Vol(S^4) \; \int d^4 x \; \int dz \; \frac{2}{3} g_s u_{KK}^5
\left( \frac{R}{u_{KK}} \right )^{\frac{3}{2}}
(1+z^2)^{\frac{2}{3}} \\
&& \left. \right. \hspace{4cm}\times \sqrt{1+\frac{9}{4 u_{KK}^2}
\left ( \frac{u_{KK}}{R} \right )^3 z^2(1+z^2)^{\frac{1}{3}}
\tau'(z)^2} \end{array}\eeq

One finds the extremal configurations $\tau(z)$ for the $D8$ obey

\begin{equation}
\tau'(z)=\frac{2}{3} \left ( \frac{R}{u_{KK}} \right )^{\frac{3}{2}} \frac{J}{\sqrt{u_{KK}^6 g_s^2 z^4(1+z^2)^2-J^2
u_{KK}^{-2} z^2(1+z^2)^{\frac{1}{3}}}}
\end{equation}

Here $J=g_s u_{KK}^4 z_0(1+z_0^2)^{\frac{5}{6}}$ is chosen effectively to make
the  gradient infinite at $z=z_0$.  This point is the point of
closest approach of the $D8$ to the horizon at $u=u_{KK}$.

This gives us a one-parameter family of embeddings where choosing
a particular value of $z_0$ specifies one particular curve.  Some
examples are shown in Figure 2 for $z_0$ increasing in factors of
$\sqrt{10}$.  Note the curve for $z_0=0$ consists of two
horizontal pieces at $\tau= \pm \frac{\pi}{3} \frac{R^{\frac{3}{2}}}{u_{KK}^{\frac{1}{2}}}$ plus a vertical
piece at $z=0$ connecting the two.  The vertical piece lies on the
horizon.

\begin{centering}
\begin{figure*}
\begin{centering}
\includegraphics[width=80mm]{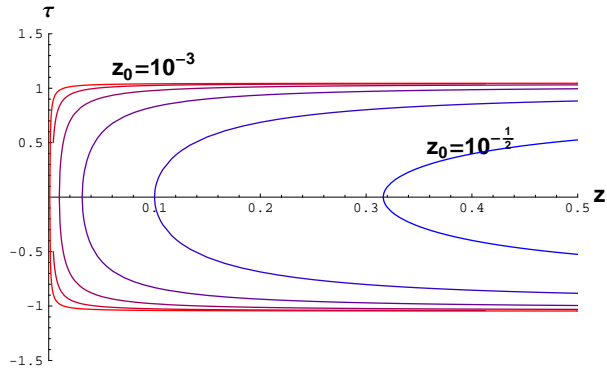}
\par
\end{centering}
\caption{Some regular $D8$-brane embeddings in the $z,\tau$-plane.
We have set $R=1$ and $u_{KK}=1$ for the numerical plot.}
\end{figure*}
\par
\end{centering}

The large $z$ (UV) asymptotic behaviour of the solutions takes the
form

\beq  \label{asymptotic} \tau = {c} - {m \over z^3} \eeq

We can determine the parameter $m$ in terms of the value of $z$
where the $D8$ and $\bar{D8}$ join ($z=z_0$) as follows. At large
$z$ the expression for $\tau'(z)$ becomes

\beq \tau'(z)= + \frac{2}{3} \frac{R^{\frac{3}{2}}}{u_{KK}^{\frac{9}{2}} g_s} \frac{J}{z^4} \eeq

This can be integrated to give

\beq \tau(z) = c - \frac{2}{9} \frac{R^{\frac{3}{2}}}{u_{KK}^{\frac{9}{2}} g_s} \frac{J}{z^3} \eeq

$c$ is the constant and one can see $m=\frac{2}{9} \frac{R^{\frac{3}{2}}}{u_{KK}^{\frac{9}{2}} g_s} \;
 J \equiv \frac{2}{9} \frac{R^{\frac{3}{2}}}{u_{KK}^{\frac{1}{2}}} z_0 (1+z_0^2)^{\frac{5}{6}}$.\\

For any given value of $c$ there is a unique regular embedding
which fixes the parameter $m$ (or equally for each value of $m$
there is a unique value of $c$).

We have argued in Section 2, from the perturbative brane construction, that a 
mass is present for the non-anti-podal embeddings. Simply from the solutions
it appears that the parameter $m$ measures the quark mass. This parameter is
zero for the embedding with the $D8$ and $\bar{D8}$ at antipodal
points on the $S^1$. It is infinite as $z_0 \rightarrow \infty$
and the quarks are removed from the theory by the $D8$ and
$\bar{D8}$ lying on top of each other.

Equally the parameter $c$ appears to measure the quark
condensate (it is zero as the mass goes to infinity and largest at
$m=0$).

This identification is not so straightforward though. Fluctuations of the 
D8 branes contain information about operators of the form
$\bar{q}_L q_R$ from strings stretched between the D8 and $\bar{D8}$ but
also operators in the adjoint of the left (or right) handed groups from
strings with both ends on one brane. The asymptotic fluctuations therefore
are most naturally associated with a coupling and vev for the operators
$\bar{q}_L D^\mu \gamma_\mu q_L$ and $\bar{q}_R D^\mu \gamma_\mu q_R$
\cite{Antonyan:2006vw}. 
However, the linking of the D8 and $\bar{D8}$ brane show that these vevs are
linked to the condensate and mass term for the quarks. Indeed in a 
gauge theory one would expect all dynamically generated coupings and masses
to be determined by the bare Lagrangian quark's mass. The identification
of $m$ and $c$ with the quark condensate and mass is thus indirect but
they do nevertheless provide a measure of those quantities. We propose that
this is why the mass shows up here as the normalizable mode and the condensate
as the non-normalizable mode. 

That these configurations decribe the gauge theory at non-zero mass
is what we would expect from the usual
holographic arguments that if the embedding contains any
information about the condensate (which it does because it
provides the effective quark mass in the theory via the $D8$-$D4$
separation) it must also describe a quark mass. The reason is that
if we write the mass terms in the field theory action as

\beq S = \int d^4x ~( m \bar{q}_L q_R + m^* \bar{q}_R q_L) \eeq
then, requiring the action to be symmetry invariant, we see that
$m^*$ and $\bar{q}_L q_R$ have the same global symmetry charges.
The holographic field associated with the condensate operator
shares these charges (there should be a unique operator-field map
if the correspondence makes sense) and so necessarily one can't
distinguish $m^*$ and $\bar{q}_L q_R$. The presence of two
constants in the solutions of the second order equation of motion
also naturally match the double role. It would therefore be very surprising
if a mass were not present in the configurations - if it weren't one would 
have to search for a larger set of D8 embeddings, but there are no such
extra embeddings.

We plot $c$ which we take as a measure of the quark condensate as
a function of the quark mass parameter $m$ in Figure 3 - it takes
a numerical value of $\frac{\pi}{3} \frac{R^{\frac{3}{2}}}{u_{KK}^{\frac{1}{2}}}$ for zero quark mass and
decreases monotonically as we increase the quark mass.

\begin{centering}
\begin{figure*}
\begin{centering}
\includegraphics[width=100mm]{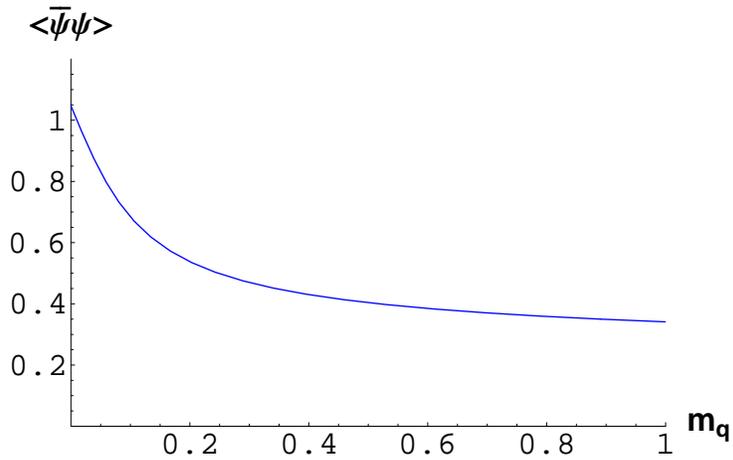}
\par
\end{centering}
\caption{`Chiral condensate' as a function of quark mass - the
parameter $c$ plotted against $m$ (with $R=u_{KK}=1$). }
\end{figure*}
\par
\end{centering}

\subsection{Vacuum Manifold and Pions in the Massless Limit}

Let us first of all review the origin of the pion in the
construction with the embedding that brings the $D8$ branes to
$z=0$. The branes lie flat in this limit at antipodal points on
the $\tau$ circle down to $z=0$.

If chiral symmetry is broken there should be a vacuum manifold
with different points corresponding to the different possible
phases on the quark condensate. The embedding function $\tau(z)$
is a real number so how do we place a phase on the parameter $c$
in its asymptotic solution? In \cite{Sakai:2004cn} the phase was
identified with the value of the gauge field $A_z$ living on the
$D8$ world volume. There is therefore a complex holographic field
that describes the condensate

\beq \label{id1} \Phi = \tau(z) e^{i A_z} = |\langle \bar{q}_L q_R
\rangle| e^{i \pi} \eeq Note here that by $\langle \bar{q}_L q_R
\rangle$ we mean a holographic function of $z$ that describes the
condensate.

To identify the vacuum manifold we should find background
solutions (that is, independent of the $x_4$ co-ordinates) for
$A_z(z,x_4)$ which correspond to different global choices of the
phase $\pi$. $A_z$ is described by the DBI action including a U(1)
gauge field, which at low energy has the Lagrangian density on the
D8 world-volume

\beq {\cal L}= e^{-\phi} \sqrt{- Det[\mathcal{P}\left ( g_{ab}
\right )]} \; \left ( -1-\frac{1}{4} F^{ab} F_{ab} \right ) \eeq

For the massless D8-brane embedding we can take $\tau(z) = \pm
\frac{\delta \tau}{4}$ which evaluates to $\pm \frac{\pi}{3} \frac{R^{\frac{3}{2}}}{u_{KK}^{\frac{1}{2}}}$ .  Physically, the vertical part of
the D8-brane in this case can be neglected because it lies along
the horizon where points separated in $\tau$ are degenerate.
Working on the upper branch of the D8-brane
($\tau(z)=+\frac{\pi}{3} \frac{R^{\frac{3}{2}}}{u_{KK}^{\frac{1}{2}}}$) the action then takes the simple form
(neglecting the volume factor coming from the four-sphere angular
coordinates - we are working with states of zero spin on the
$S^4$ here)

\begin{equation} \label{gaugeaction}
{\cal S} = \frac{1}{2} \int_0^{\infty} dz \int d^4 x \; \left  (
e^{-\phi} \sqrt{-g} g^{zz} g^{11} \right ) \; \left (-(\partial_0
A_z)^2+(\partial_1 A_z)^2+(\partial_2 A_z)^2+(\partial_3 A_z)^2
\right )
\end{equation}

More explicitly this is

\begin{equation}
{\cal S} = \frac{3}{4} g_s u_{KK}^3 \left ( \frac{R}{u_{KK}}
\right )^{\frac{3}{2}} \int_0^{\infty} dz \int d^4 x \; (1+z^2) \;
\left (-(\partial_0 A_z)^2+(\partial_1 A_z)^2+(\partial_2
A_z)^2+(\partial_3 A_z)^2 \right )
\end{equation}

It is apparent that $F^{ab}$ and hence the action vanishes if
$A_z$ is the only non-zero field and if it is only a function of
$z$. Any function of $z$ is allowed. This is an artifact of gauge
freedom in the model and one should pick a gauge. For example one
could gauge fix by including a term

\beq \delta {\cal L} = {1 \over \xi} \; e^{-\phi} \sqrt{-
Det[\mathcal{P}\left ( g_{ab} \right )]} \; \left ( \nabla_a A^a -
\kappa(z)\right )^2  \eeq where $\kappa(z)$ is any arbitrary
function. Writing $A_z(z,x_4) \equiv g(z) \pi(x_4)$, there is
sufficient freedom to pick any functional form of $g(z)$. We will
follow the choice of Sakai and Sugimoto and pick

\beq g(z) = \frac{\rm \cal{C}}{1+z^2}  \eeq

The solution contains the arbitrary multiplicative factor ${\cal
C}$ since the action is only quadratic in $A^z$. The freedom to
pick the constant ${\cal C}$ in this solution is the freedom to
move on the vacuum manifold.

We can now identify the pion field. It should correspond to
space-time ($x^\mu$) dependent fluctuations around the vacuum
manifold. In other words we look at solutions of the form

\beq A_z(z,x) = \pi(x_4) \times \frac{2}{\sqrt{3 \pi}} \;
\frac{1}{1+z^2} \eeq

Substituting this into the action (13) we find a canonically
normalized kinetic term for a massless field.

\beq {\cal S} = \int d^4 x {1 \over 2} (\partial^\mu \pi)^2 \eeq

This is the pion - the Goldstone mode of the chiral symmetry
breaking (although we call it a pion in QCD it is the $\eta'$
which is not a Goldstone due to anomaly arguments - at large $N$
this field is closer in spirit to the pions of QCD).

\section{Massive Embeddings}

The analysis of $A_z$ in the case of the embeddings that we claim
also describe a quark mass appears initially to be identical to
the massless case above. We can define, using the gauge freedom, a
background configuration where

\beq A_z = {\rm constant} \times {1 \over \sqrt{e^{-\phi}\sqrt{-g}
g^{zz} g^{11}}} \frac{1}{\sqrt{1+z^2}} \eeq where the constant is
chosen to make the coefficient of the kinetic term  equal to
$\frac{1}{2}$ when we integrate over $z$ to give a 4D action (as
is apparent from (\ref{gaugeaction}))\footnote{Note that naively
$A_z$ is singular in these ansaetze but the singularity at the
point of closest approach $z_0$ in each case is a result of the
singularity in the embedding function $\tau(z)$. The D8 embedding
is not really singular at these points though. The singularity
results from using the coordinate $z$ as the coordinate on the D8
world volume - this is clearly inappropriate at $z_0$. This in
fact is also the case in the massless limit although there the
embedding only deviates from a flat embedding at precisely the
point $z_0$. One could work with $\tau$ as the coordinate and
express the embedding through a function $z(\tau)$ and then $z_0$
would be singularity free but there would be a singularity at
large $z$. We will live with the coordinate induced singularity in
our equations. The singularity also makes issues of whether
$\pi(z)$ is an even or odd function tricky - in fact though it is
even since the embedding function is (this is necessary for the
pion to be a pseudo-scalar - see \cite{Sakai:2004cn}). The even
property is again most easily seen by interchanging the roles of
$\tau$ and $z$.}.

We can again find a space-time dependent field

\beq A_z = \pi(x_4) \times {1 \over \sqrt{e^{-\phi}\sqrt{-g}
g^{zz} g^{11}}} \frac{1}{\sqrt{1+z^2}}  \eeq which on substitution
back into the action appears to give a massless state. Is there
therefore a Goldstone even for these embeddings hence invalidating
our interpretation of there being a quark mass present?

In fact we have been cavalier. We must be careful to correctly
identify what this massless field is. In particular we need an
equation equivalent to (\ref{id1}) above telling us what $A_z$ is
holographically describing. Led by (\ref{id1}) the natural choice
is

\beq \label{id2} \Phi = \tau(z) e^{i A_z} = \langle \bar{q}_L q_R
\rangle + m^* \eeq We again mean on the right here holographic
functions of $z$ that describe the condensate and mass.

It is now immediately clear that changing $A_z$ corresponds to
changing the phase of both the quark condensate and the quark mass
together. This is a spurious transformation that is indeed a flat
direction in the potential in the space of such theories and the
presence of the vacuum degeneracy and the `Goldstone' is now seen
to be natural. To be able to make this transformation though we
must be able to change the mass parameter of the theory moving us
to a different theory. It is not an allowed transformation with in
any one theory. This is not a physical state.

It is worth noting that a holographic field of precisely the form
in (\ref{id2}) is found in the chiral symmetry breaking models
using D3/D7 \cite{Babington:2003vm} and D4/D6 systems
\cite{Kruczenski:2003uq}. As we argue is the case here, those
models have a continuous set of degenerate probe configurations
even when the quark mass is non-zero - it is precisely the
spurious symmetry identified above that moves between these
configurations (in those models it is an explicit rotational
symmetry of a spacetime plane). The presence of the spurious
symmetry that transforms both the mass and the condensate is to be
expected because one has the ability to determine the mass through
the holographic field. This symmetry ought to be present in the
Sakai-Sugimoto model and our interpretation correctly explains it.

To identify the true pion in the Sakai-Sugimoto model we must
change the phase on the quark condensate whilst leaving the phase
on the mass unchanged. It is clear from (\ref{id2}) that switching
on the pion in this way will force a fluctuation in the embedding
function $\tau(z)$ - since there is no flat direction for this
embedding we expect the pion to acquire a mass. We write

\begin{equation}
\Phi = \langle \bar{q}_L q_R \rangle e^{i \pi(z,x_4)} + m^*
\end{equation}
and rewrite this in modulus-argument form and expand to quadratic
order in the pion field. This tells us via (\ref{tapi}) how the
embedding and the $A_z$ field are perturbed by the pion. The
result is

\begin{equation} \label{tapi}
\tau(z,x_4) = \langle \bar{q}_L q_R \rangle  + m^* - \frac{\langle
\bar{q}_L q_R \rangle  m^* \pi(z,x_4)^2}{2 (\langle \bar{q}_L q_R
\rangle  + m^*)}, \hspace{1cm} A_z(z,x_4) = \frac{ \langle
\bar{q}_L q_R \rangle \pi(z,x_4)}{\langle \bar{q}_L q_R \rangle
 + m^*}
\end{equation}
Note that $A_z=\pi(z,x_4)$ in the massless case.

At this point one would like to identify which part of the
background embedding function, which we will hence forth call
$\tau_0(z)$, represents the $z$ dependence (renormalization group
flow) of the condensate and which represents the $z$ dependence of
the mass. In general there is no obvious way to make this split (a
similar problem exists in other holographic models of chiral
symmetry breaking as discussed in \cite{Evans:2004ia}). Let us
simply write $\tau_0=c(z)+m(z)$ where $c(z)$ and $m(z)$ are
functions that asymptote to the two terms in (\ref{asymptotic}).

To quadratic order in the pion field we have

\begin{equation}
\tau(z,x_4) = \tau_0 - \frac{c(z) m(z) \pi(z,x_4)^2}{2 \tau_0},
\hspace{1cm} A_z(z,x_4) = \frac{ c(z) \pi(z,x_4)}{\tau_0}
\end{equation}

These describe perturbations to the embedding and gauge field such
that substituting the above into the DBI action will give a term
of order zero in $\pi$ which will recover the background embedding
functions $\tau_0(z)$ and a term quadratic in $\pi$ giving linear
dynamics to our pion field.

As before  our strategy will be to write down a 5D action and use
the gauge freedom to reduce the kinetic term to canonical form
when integrated over $z$ to give a 4D pion action.  This time, for
a massive embedding we will have an extra term in the 4D action
(from the curvature of the embedding) which will generate a mass
squared term.

The kinetic term in the 4D Lagrangian from the DBI action for the
gauge field is now (separating variables so $\pi(z,x_4) \equiv
g(z) \pi(x_4)$)

\begin{equation}
{\cal L}_{kinetic}= \int dz \frac{3}{4} g_s u_{KK}^3  \left(
\frac{R}{u_{KK}} \right )^{\frac{3}{2}} \left (
\frac{c(z)}{\tau_0(z)} \right )^2 \frac{(1+z^2) g(z)^2
\partial_{\mu} \pi(x_4) \partial^{\mu} \pi(x_4)}
{\sqrt{1+\frac{9}{4 u_{KK}^2} \left ( \frac{u_{KK}}{R} \right )^3
z^2 (1+z^2)^{\frac{1}{3}} \tau_0'(z)^2}}
\end{equation}

To normalize our 4D kinetic term canonically we should choose

\begin{equation}
g(z)^2 = \frac{4}{3} \frac{1}{g_s u_{KK}^3 \left(
\frac{R}{u_{KK}} \right )^3}\left ( \frac{\tau_0(z)}{c(z)} \right
)^2 \frac{\sqrt{1+\frac{9}{4 u_{KK}^2} \left ( \frac{u_{KK}}{R}
\right )^3 z^2(1+z^2)^{\frac{1}{3}} \tau_0'(z)^2}}{1+z^2}
\frac{1}{1+z^2} \frac{1}{2 \int_{z_0}^{\infty} \frac{dx}{1+x^2}}
\end{equation}

For the massive embedding there is though an additional term in
the Lagrangian coming from the change in the embedding in
(\ref{tapi}) above. We find, expanding $e^{-\phi} \sqrt{-g}$ which
is (\ref{taction}) with the perturbation switched on (so we
replace $\tau(z)$ in (\ref{taction}) with
$\tau_0(z)-\frac{c(z)m(z) \pi(z,x_4)^2}{2\tau_0}$ and expand the
radical as $\sqrt{a+b} \sim \sqrt{a}+\frac{b}{2 \sqrt{a}}$)

\begin{equation}
{\cal L}_{mass}=\int dz \frac{3}{4} g_s u_{KK}^3 \left  (
\frac{u_{KK}}{R} \right )^{\frac{3}{2}} \; \frac{z^2(1+z^2)
\tau_0'(z) \left ( \frac{m(z) c(z) g(z)^2}{\tau_0(z)} \right
)'}{\sqrt{1+\frac{9}{4 u_{KK}^2} \left ( \frac{u_{KK}}{R} \right
)^3 z^2(1+z^2)^{\frac{1}{3}} \tau_0'(z)^2}} \; \pi(x_4)^2
\end{equation}

One can see that for the case of massless quarks, where
$\tau'_0(z)=0$, the expression correctly gives a vanishing pion
mass. Away from that limit it will be non-zero and the pion
becomes a pseudo-Goldstone boson. Unfortunately any explicit
computation requires a deeper understanding of the split of
$\tau_0$ into the parts $m(z)$ and $c(z)$. Nevertheless this
computation provides support for our interpretation of the
embeddings given above.

\section{Summary}

We have re-analyzed the family of embeddings of the Sakai-Sugimoto
holographic model of chiral symmetry breaking. We have argued that
the natural interpretation of the asymptotic parameters is as the
quark mass and condensate - this matches the perturbative string
picture and the behaviour of the effective quark mass in the
model. We have shown that in this identification a previous
argument that all the embeddings have a massless pion is incorrect
- the massless state is a spurious field corresponding to allowing
the quark mass to change it's phase. The true pion state develops
a non-zero mass on these embeddings consistent with the
interpretation that an explicit quark mass is present.
\bigskip

\noindent {\bf Acknowledgements:} The authors would like to thank
David Mateos and Shigeki Sugimoto for discussions. Ed Threlfall is
grateful to PPARC for the support of a studentship.

\end{document}